\documentclass[10pt]{article}
\usepackage{textgreek}
\usepackage{xfrac}
\usepackage{amsmath}
\usepackage{array}

\begin{document}
\begin{Large}
\noindent\textbf{Visible spectrum extended-focus optical coherence microscopy for label-free sub-cellular tomography\\
}
\end{Large}

\noindent Paul J. Marchand,$^{1,*}$ Arno Bouwens,$^{1}$ Daniel Szlag,$^{1}$ 
David Nguyen,$^{1}$ Adrien Descloux,$^{1}$ Miguel Sison,$^{1}$ S\'{e}verine Coquoz,$^{1}$ J\'{e}r\^{o}me Extermann,$^{1}$ and Theo Lasser$^{1}$\\

\noindent$^{1}$Laboratoire d\textquoteright Optique Biom\'{e}dicale, Ecole Polytechnique F\'{e}d\'{e}rale de Lausanne, CH-1015 Lausanne, Switzerland\\

$^{*}$paul.marchand@epfl.ch 



\begin{abstract}
We present a novel extended-focus optical coherence microscope (OCM) attaining 0.7 \textmu m axial and 0.4 \textmu m lateral resolution maintained over a depth of 40 \textmu m, while preserving the advantages of Fourier domain OCM. Our method uses an ultra-broad spectrum from a super-continuum laser source. As the spectrum spans from near-infrared to visible wavelengths (240 nm in bandwidth), we call the method visOCM. The combination of such a broad spectrum with a high-NA objective creates an almost isotropic 3D submicron resolution. We analyze the imaging performance of visOCM on microbead samples and demonstrate its image quality on cell cultures and ex-vivo mouse brain tissue.
\end{abstract}


\section{Introduction}

Over the past decades, optical microscopy has allowed investigating biological systems at high spatial and temporal resolution. Confocal fluorescence microscopy \cite{Pawley2006} and light-sheet microscopy\cite{Huisken1007}, through their capabilities in three-dimensional imaging, have become the mainstay for cellular and subcellular imaging. Nevertheless, while fluorescence provides molecular specificity, the influence of these agents on cellular processes is ambiguous as they might interfere with the functioning of the cell. These effects combined with photobleaching ultimately hinder the possibility to perform long-term imaging. 

In such studies, label-free microscopy offers an interesting alternative as it can provide wide-field images at high acquisition rates without using exogenous agents. Moreover, the absence of labels facilitates the sample preparation. Recent advances in phase microscopy \cite{Kim2014} and ptychography \cite{Tian2015} have allowed performing three-dimensional imaging of cellular cultures or embryos but remain limited to thin single layer structures. %

Optical coherence tomography (OCT) is an interferometric imaging technique sensitive to refractive index contrast in the sample \cite{Fercher2003}. In OCT, the axial resolution is defined by the width of the illumination spectrum and an entire depth profile can be obtained from a single recording of the output spectrum. As such, only a two-dimensional scan is required to obtain a three-dimensional image.

Optical coherence microscopy (OCM), the microscopy analogue to OCT, uses high-NA objectives to obtain a higher lateral resolution. In standard OCM systems, however, the axial field of view is dictated by the Rayleigh range and thus decreases quadratically (\mbox{$\propto 1/$ NA$^2$}) with the improvement in lateral resolution (\mbox{$\propto 1/$ NA}). This compromise can be circumvented by engineering an extended-focus illumination through the use of so-called diffraction-less beams such as Bessel beams \cite{Leitgeb2006}. 

In order to maintain a good collection efficiency of the scattered light signal, a Gaussian detection mode is used. Therefore, separate illumination and detection modes are required: a Bessel illumination mode and a Gaussian detection mode. This split between modes can further be exploited to filter specular reflections and obtain a dark-field OCM system \cite{Villiger2010}. The dark-field property is particularly important when investigating weakly scattering structures, such as cell samples, as it suppresses light reflected from the sample support which would otherwise strongly reduce the usable dynamic range of the detector. As such, all available dynamic range can be devoted to the desired, but weak, scattered light signal.

In this paper, we present visible spectrum optical coherence microscopy (visOCM). The system builds upon our previous dark-field OCM design, and improves its imaging capabilities for sub-cellular structures by using a large bandwidth illumination spectrum spanning visible to near-infrared wavelengths and a high-NA objective. The resulting system possesses an almost isotropic submicron resolution (0.4 \textmu m laterally and 0.7 \textmu m axially) maintained over a large depth of field (>40 \textmu m). Hence  visOCM extends the capabilities of our previous non-imaging visible light optical coherence correlation spectroscopy (OCCS) system and is optimized three-dimensional cellular tomography \cite{Broillet2014}. We present a strategy for dispersion compensation and demonstrate the system's 3D resolution on microbead samples. We demonstrate visOCM's image quality and contrast on living cell cultures as well as fixed brain slices of healthy and alzheimeric mice.

Besides imaging the structure of a sample at a given time-point, there is also much interest in monitoring intracellular dynamics to understanding cell function. As such, several optical microscopy techniques have been developed to analyse cell trafficking and intracellular motility\cite{Sison2017,Oldenburg2015,Ma2016}. Recently, OCT methods developed to obtain qualitative and quantitative information on vascular function have been used to reveal sub-cellular compartments and quantify their activity \cite{Lee2013,Apelian2016}. Being a Fourier-domain method, visOCM is capable of rapidly acquiring tomograms, and therefore these dynamic signal imaging methods can be applied to visOCM as well. We demonstrate dynamic signal imaging with visOCM on living cells.

\section{Materials and Methods}
\subsection{Optical Setup}
\begin{figure}[htbp]
\centering\includegraphics[width=13cm]{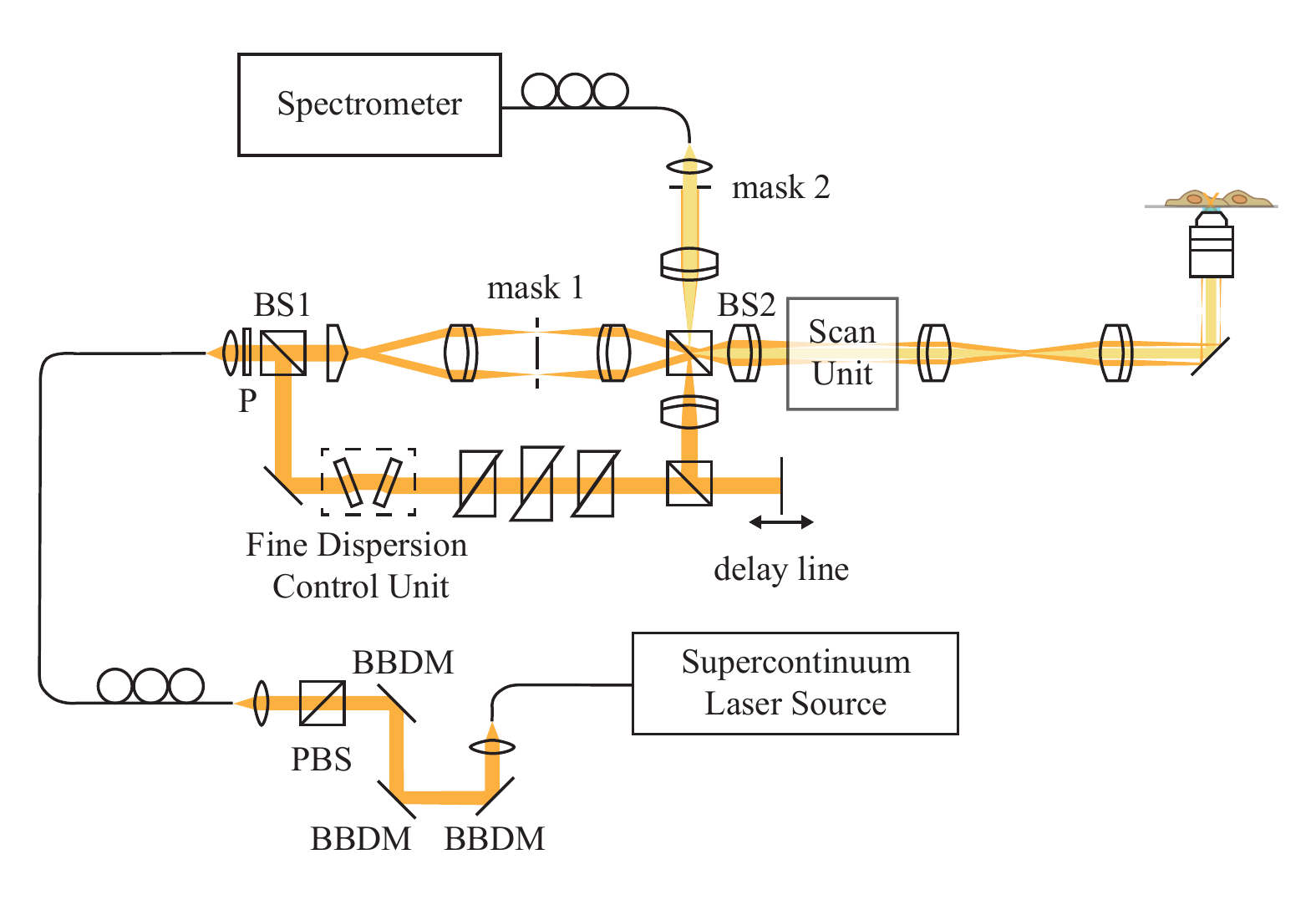}
\caption{Schematic of the extended-focus OCM using a broad spectrum in the visible wavelength range and a high NA objective for high axial and lateral resolution. By combining a Bessel illumination, generated by an axicon lens, and a Gaussian detection, a dark-field extended-focus system can be obtained. P: Polarizer, BBDM: Broadband dielectric mirror, BS: Beamsplitter, PBS: Polarizing beamsplitter.}
\label{Fig:setup}
\end{figure}

As illustrated in Figure \ref{Fig:setup}, the optical setup is based on a Mach-Zehnder interferometer, allowing for a separation of the illumination and detection modes, necessary to obtain the desired Bessel-Gauss configuration. The output of a supercontinuum laser (Koheras SuperK Extreme, NKT Photonics) is first filtered through three broadband dielectric mirrors (BB1-EO2, Thorlabs) to reject the source's strong infrared emission. Then, the light is split by a polarizing beam-splitter (PBS251, Thorlabs), and injected in the interferometer using a single-mode fiber (P3-460B-FC-2, Thorlabs). As shown in Figure \ref{Fig:charac}(a), the spectrum in the interferometer is centred at 647 nm and has a 246 nm bandwidth. After collimation, the light passes through a polarizer and then a first beamsplitter (BS1) splits the light into the reference and illumination paths, where an axicon lens (Asphericon, apex angle 176$^{\circ}$) generates a Bessel beam. The beam is then Fourier filtered in a telescope to remove stray light originating from the axicon's tip, after which it is steered to the scan-unit, and focused on the sample by a 40x high NA objective (Olympus, effective NA = 0.76). The back-scattered light is collected by the objective, de-scanned by the scan-unit and directed to the detection arm through a second beam splitter (BS2) where it is coupled to a custom-made spectrometer with a single-mode fiber (P3-460B-FC-2, Thorlabs). The spectrometer is comprised of a transmission grating (600 lines/mm, Wasatch Photonics) and a fast line-scan camera (Basler spL2048-140km). Dispersion matching between both interferometer arms was performed by adding a combination of prism pairs in the reference arm. The detailed dispersion matching strategy is presented in \ref{section:dispMatch}.

To obtain a dark-field configuration, the specular reflection was suppressed by spatially filtering the reflected Bessel ring through a mask in the detection arm (mask 2) and by properly filtering the stray light from the tip of the axicon lens (mask 1).

\subsection{Dispersion compensation strategies}
\label{section:dispMatch}
The use of a broad illumination spectrum in visOCM, renders dispersion matching more challenging, particularly in the visible spectrum where the relationship between the refractive index and the wavelength becomes increasingly non-linear at shorter wavelengths. Here we opted for a physical dispersion compensation scheme where we first estimated the dispersion caused by the illumination and sample arms prior to constructing the optical system. This allowed us to calculate the best composition of prisms to place in the reference arm (Figure \ref{Fig:disp}(a)). In a second step, during the construction of the setup, we developed a real-time interface in LabView to display the amount of residual dispersion and used a motorized stage to finely adjust the thickness of the different glasses (Figure \ref{Fig:disp}(b-c)).

\subsubsection{Estimation of the dispersion in the optical system}
Prior to constructing the optical system, we first estimated the thickness of each glass type present in the different arms of the interferometer. The glass type and respective thickness of each optical element (lenses and beamsplitters) were obtained from the data-sheets provided with each element by Thorlabs. The objective was not modelled in this initial assessment as its composition is unavailable.
As depicted in Figure \ref{Fig:disp}(a) and described in Equation \ref{Eq:OPD}, we then proceeded to estimate the \textit{k}-dependent optical path length of each part of the system (sample arm, illumination arm and reference arm) by using the thickness of each glass and their respective dispersion curves (Sellmeier coefficients from the SCHOTT database available on https://www.refractiveindex.info) to obtain the optical path difference (OPD). 

\begin{multline}
OPD(k)=
\begin{bmatrix}
n_{sample,SF5}(k),  \dots , n_{sample,UVFS}(k)
\end{bmatrix} 
\begin{bmatrix}
d_{sample,SF5}\\
\vdots\\
d_{sample,UVFS}\\
\end{bmatrix} \dots\\
+\begin{bmatrix}
n_{ill,SF5}(k), \dots, n_{ill,UVFS}(k)
\end{bmatrix} 
\begin{bmatrix}
d_{ill,SF5}\\
\vdots\\
d_{ill,UVFS}\\
\end{bmatrix} \dots\\
-\begin{bmatrix}
n_{ref,SF5}(k),  \dots , n_{ref,UVFS}(k)
\end{bmatrix} 
\begin{bmatrix}
d_{ref,SF5}\\
\vdots\\
d_{ref,UVFS}\\
\end{bmatrix}
\label{Eq:OPD}
\end{multline}

The obtained OPD was then fit to a set of dispersion curves through a multivariate linear regression to find the thickness of the set of glasses to best balance the dispersion. As described in Equation \ref{Eq:MVR}, the multivariate linear regression tries to model the OPD as the sum of the refractive indexes (in function of \textit{k}) of each glass multiplied by their respective thicknesses and an error term $\epsilon(k)$. By performing this analysis, we can therefore retrieve the thicknesses of each glass required to compensate the OPD of the optical system. An initial analysis of several sets of glasses revealed that combining BK7 and SF10 allowed fully balancing the dispersion of the system. 

\begin{equation}
OPD(k)=
\begin{bmatrix}
n_{BK7}(k), n_{SF10}(k)
\end{bmatrix} 
\begin{bmatrix}
d_{BK7}\\
d_{SF10}\\
\end{bmatrix}
+\epsilon(k)
\label{Eq:MVR}
\end{equation}

\subsubsection{Fine matching of the dispersion}
\begin{figure}[htbp]
\centering\includegraphics[width=13cm]{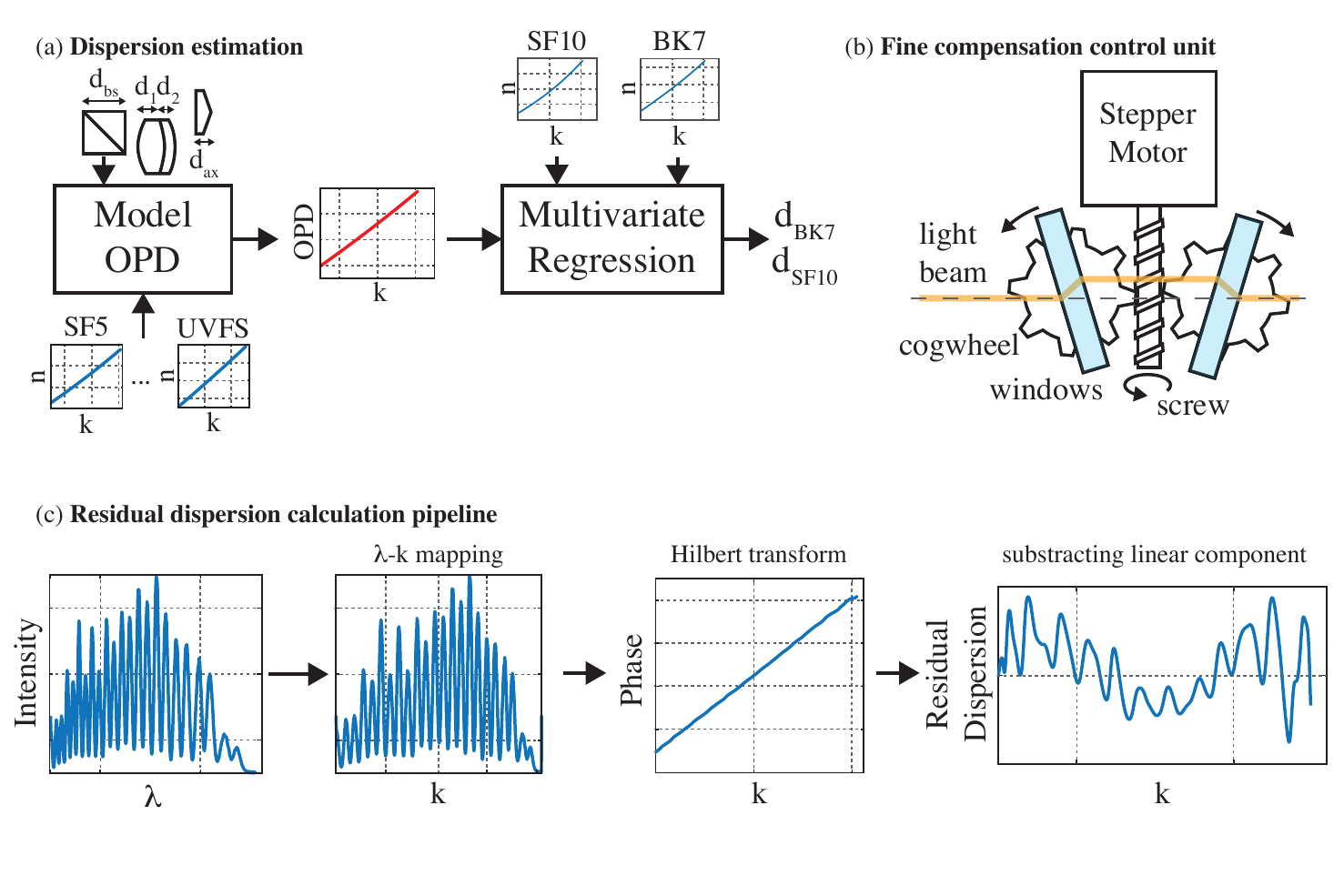}
\caption{Dispersion compensation strategies: (a) Prior to aligning the optical setup, a first estimation of the dispersion mismatch in the system is performed by modelling the optical path length of the different arms of the interferometer. The thicknesses and dispersion curves of the different glasses are used to model the OPD between the arms of the system. The OPD is then fit with a multivariate regression using the dispersion curves of SF10 and BK7 as set of regressors to obtain the respective thicknesses to fully balance the dispersion in the interferometer. (b) In order to match the dispersion, a compensation unit was devised allowing a fine control of the thickness of the glass. A mechanism comprising a stepper motor and cogwheels allows changing the angle of a pair of glass windows to vary the length of glass traversed by light. (c) During the alignment, the dispersion was matched by observing and minimizing the residual dispersion present in the system, obtained by a simple set of processing steps: The interferogram first undergoes a $\lambda$-k mapping step from which the phase is then extracted by a Hilbert transform. The linear component of this phase is then extracted and removed to reveal the residual dispersion.}
\label{Fig:disp}
\end{figure}

During the construction and alignment of the system, the dispersion was finely balanced by calculating the residual dispersion in the system and by adjusting the thickness of each individual glass through a pair of motorized mechanical stages. The residual dispersion was measured by placing a sample consisting of a cover glass and a mirror. Then the interferogram's phase was calculated by Hilbert transformation \cite{Wojtkowski2004}, and a linear fit was subtracted (Figure \ref{Fig:disp}(c)). If the dispersion is perfectly balanced between the arms of the interferometer then the residual dispersion curve should be null throughout the spectrum. 
The motorized mechanical stages, as depicted in Figure \ref{Fig:disp}(b), consists of a stepper motor operating a cogwheel mechanism. The mechanism allows varying the angle between two glass windows, in order to fine-tune the amount of that glass type in the reference arm. Moreover, as the system is symmetrical, changes in the beam's transverse position caused by the passage of the light through the first window are compensated by the second window. As such, varying the amount of glass causes minimal changes in the alignment of the optical system. The combination of these two effects (fine-tuning and minimal misalignment) facilitates the dispersion compensation procedure. By observing the residual dispersion and iteratively changing the thickness of each glass we could balance the dispersion of the two arms of the microscope. We used SF6 windows, instead of SF10, for dispersion fine tuning.

\subsection{Image acquisition and processing}
All images were acquired at a 20 kHz A-scan rate (with an integration time of 43 \textmu s) and with a power varying from 1.5 mW to 3 mW in the back-focal plane of the objective. With the exception of the dynamic signal imaging protocol (presented in \ref{sec:DynSigIm}), the size of each image was 512 $\times$ 512 $\times$ 2048 pixels (\textit{x},\textit{y} and \textit{k} respectively).
Large field-of-views were obtained by stitching several tomograms (each having a lateral field-of-view of either 60 \textmu m $\times$ 60 \textmu m, or 120 \textmu m $\times$ 120 \textmu m) with a 30\% overlap between each tile of the mosaic in both directions. Any tilt (angle with respect to the optical axis), was corrected on both axes (\textit{x-z} and \textit{y-z}) prior to stitching. The tomogram processing, tilt-correction and stitching were performed through a custom-coded MATLAB graphical interface. The tomograms presented in Figures \ref{Fig:PTLp}--\ref{Fig:cells} are displayed with the intensity in logarithmic scale for visualisation purposes.
The tomograms were convolved with a 3D Gaussian kernel (\textsigma$_{x,y}$ = 0.187 \textmu m, \textsigma$_{z}$ = 0.22 \textmu m) and were then resized to obtain an isotropic sampling using ImageJ.

\subsection{Sample preparation}
\subsubsection{Mice brain slices}
\label{sec:MethodsSlices}
All experiments were carried out in accordance to the Swiss legislation on animal experimentation (LPA and OPAn). The protocols (VD 3056 and VD3058) were approved by the cantonal veterinary authority of the canton de Vaud, Switzerland (SCAV, D\'{e}partement de la s\'{e}curit\'{e} et de l'environnement, Service de la consommation et des affaires v\'{e}t\'{e}rinaires) based on the recommendations issued by the regional ethical committee (i.e. the State Committee for animal experiments of canton de Vaud) and are in-line with the 3Rs and follow the ARRIVE guidelines.
Brain slices were obtained by perfusing transcardially B6SJL/f1 mice with PBS followed by 10\% Formalin (HT501128, Sigma-Aldrich). The mice were injected subcutaneously with Temgesic prior to the perfusion with heparinized PBS. The extracted brains were then left in 4\% PFA overnight, and then placed in a solution of 30\% glucose. Finally, the brains were cut into slices using a microtome at a thickness of $\sim$30 \textmu m and placed on a glass coverslide.
Brain slices from 5xFAD mice, a mice model of amyloid pathology, were obtained using the same protocol. The amyloid plaques were stained using a solution of Methoxy-X04 in DMSO, which was administered through two I.P. injections 24h and 2h before the perfusion, as described by J\"arhling et al. \cite{Jahrling2015}.

\subsubsection{Macrophages preparation}

In addition to mice brain slice imaging, we imaged live murine macrophages (cell line RAW 264.7) with the visOCM platform. The RAW 264.7 cells were cultured in an incubator at 37$^\circ$C and 5\% CO$_{2}$ using DMEM high glucose with pyruvate (4.5 g l-1 glucose, Roti\textregistered-CELL DMEM, Roth) supplemented with 10\% fetal bovine serum and 1\texttimes  penicillin-streptomycin (both gibco\textregistered, Thermo Fisher Scientific). Prior to imaging (1-2 days), the cells were seeded in FluoroDish Sterile Culture Dishes (35 mm, World Precision Instruments).

\section{Results}
\subsection{System Characterization}
\begin{figure}[htbp]
\centering\includegraphics[width=13cm]{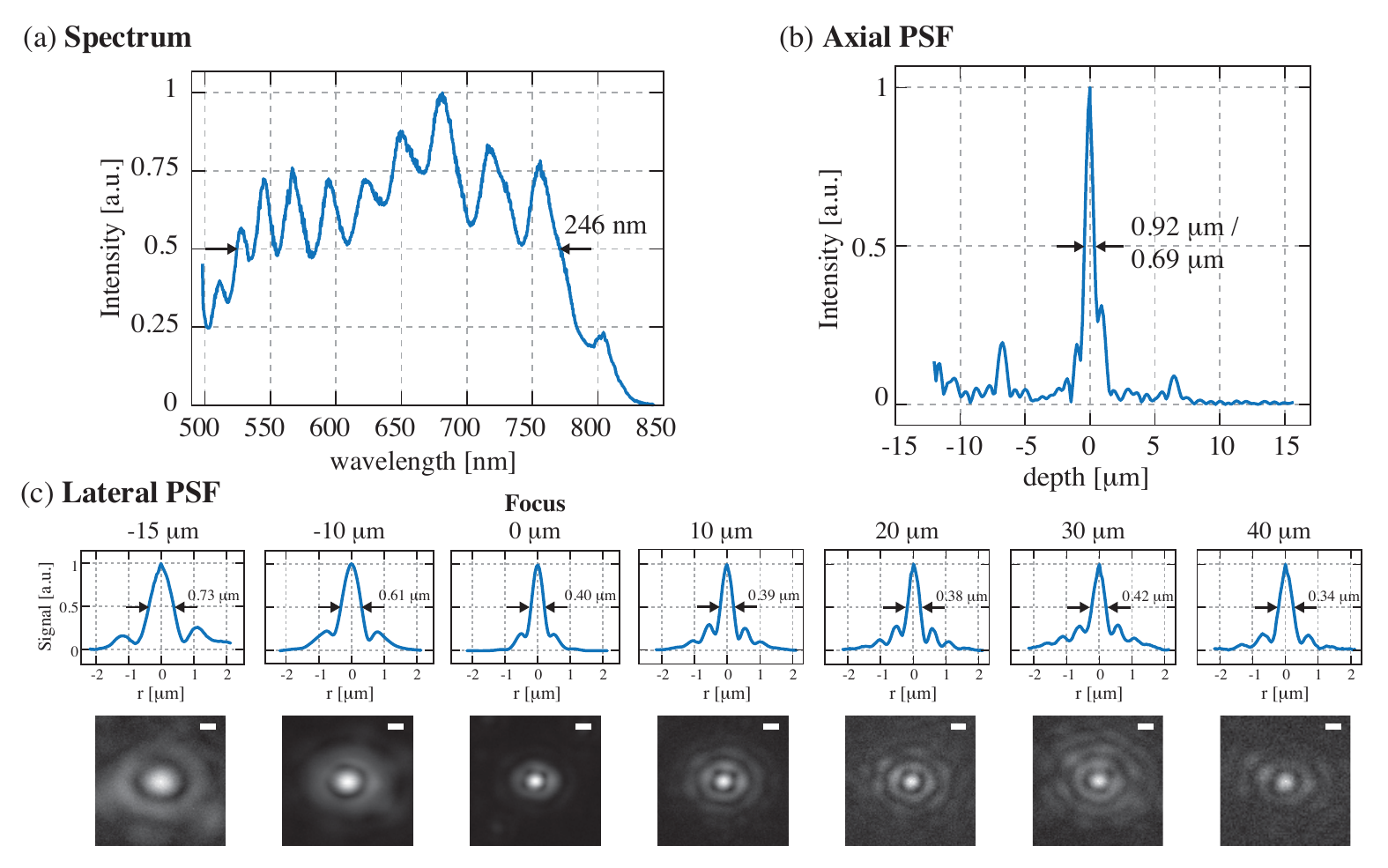}
\caption{Characterization of the visOCM system: (a) Illumination spectrum spanning from the visible to the near-infrared range, centred at 647 nm and 246 nm wide. (b)The ultra-broad spectrum leads to a submicrometric optical sectioning capability. (c) Plots and heatmaps (in linear scale) displaying the lateral PSF along the depth of focus of the objective, illustrating that the diameter of the central lobe is maintained at $\sim$400 nm over 40 $\mu$m in depth.  Scalebar: 500 nm}
\label{Fig:charac}
\end{figure}

The system's lateral resolution was characterized by imaging a solution of nanoparticles of 30 nm in diameter embedded in a slab of PDMS. The small size of these particles allowed interrogating the point-spread function (PSF) of the optical system. The depth-dependence of the lateral resolution was assessed by isolating and averaging multiple ($\sim$ 10) measurements of the PSF at 7 different depths. The lateral profile of the measured PSF at each depth was then extracted, and the position of the first zero served as a measure of the lateral resolution. As shown in Figure \ref{Fig:charac}(c) the width of the central lobe is maintained at 400 nm over 40 \textmu m.

The axial resolution of the system was measured by imaging a mirror placed on a glass coverslip in the sample arm. The reference power was adjusted to obtain maximum visibility of the interference pattern. The axial PSF, measured as mentioned previously, is plotted in Figure \ref{Fig:charac}(b) and has a FWHM of 0.92 \textmu m in air, corresponding to a width of 0.69 \textmu m in water.

\begin{figure}[htbp]
\centering\includegraphics[width=13cm]{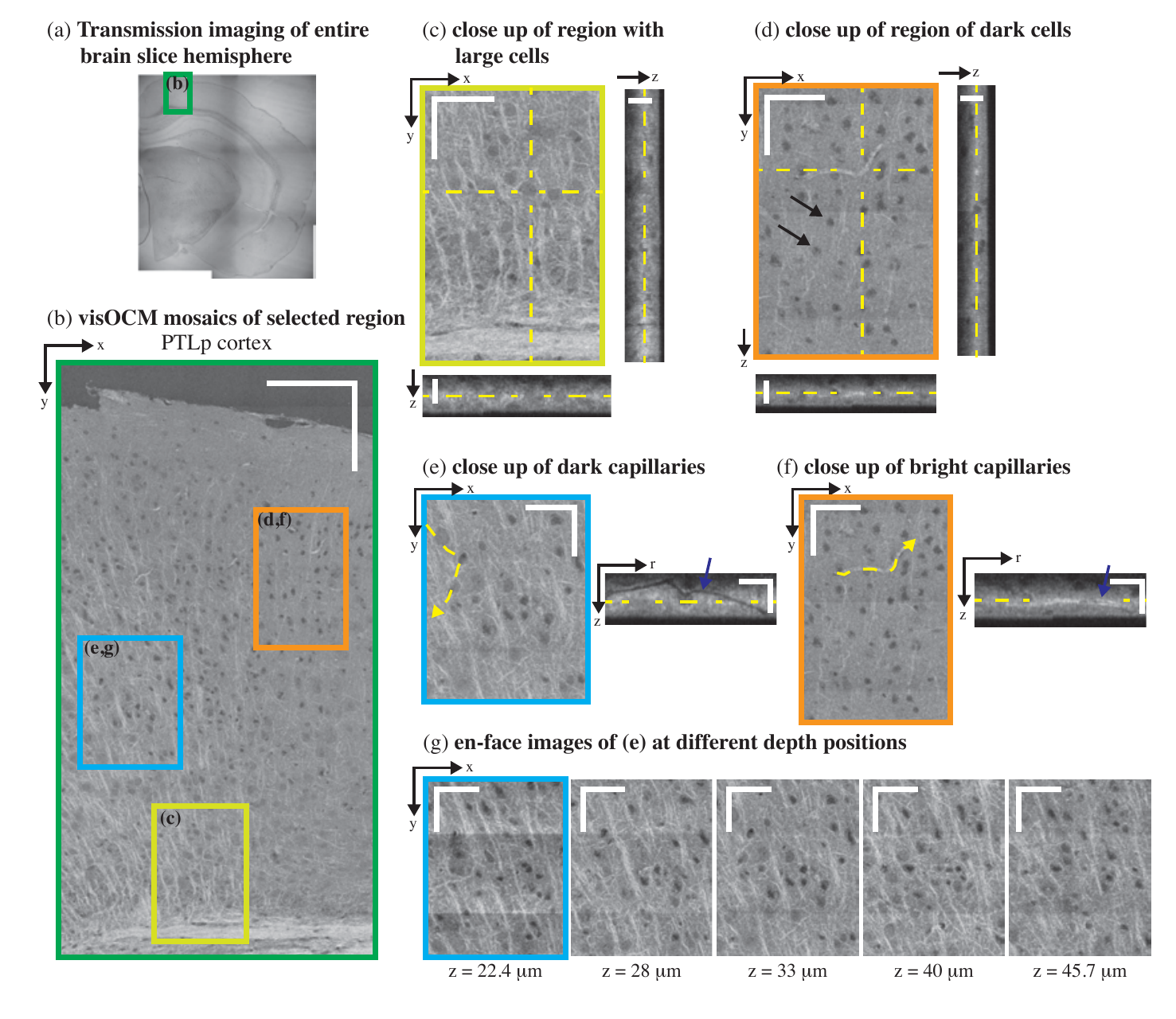}
\caption{\textit{ex-vivo} visOCM imaging of the PTLp cortex in a B6SJL/f1 mouse brain slice. (a) A transmission image of the entire mouse hemisphere was first acquired to locate the desired area (green rectangle), which was then imaged with visOCM (b). The mosaic of part of the PTLp cortex, acquired with visOCM, reveals a variety of cortical structures, such as fibers, cell bodies and vascular entities (en-face view). In the mosaic, mainly two types of cells can be visualized, large cells as shown in (c) and smaller darker cells as pointed by arrowheads in (d). The orthogonal views in (c--d) highlight the three-dimensional repartition of these cell types within the depth of the slice. Capillary vessels can be discriminated from the tissue as either dark or bright structures as shown in (e) and (f) respectively. These different contrasts are further revealed in the orthogonal slices accompanying the close-ups, where one can trace the path of the hollow dark lumen or the bright vessel border, pointed by the arrowheads. En-face images at different depths show that visOCM can perform imaging over >20 $\mu$m (g). Scalebars: 150 $\mu$m (b), 50 $\mu$m for the en-face and 20 $\mu$m for the orthogonal views (c--g).}
\label{Fig:PTLp}
\end{figure}

\begin{figure}[htbp]
\centering\includegraphics[width=13cm]{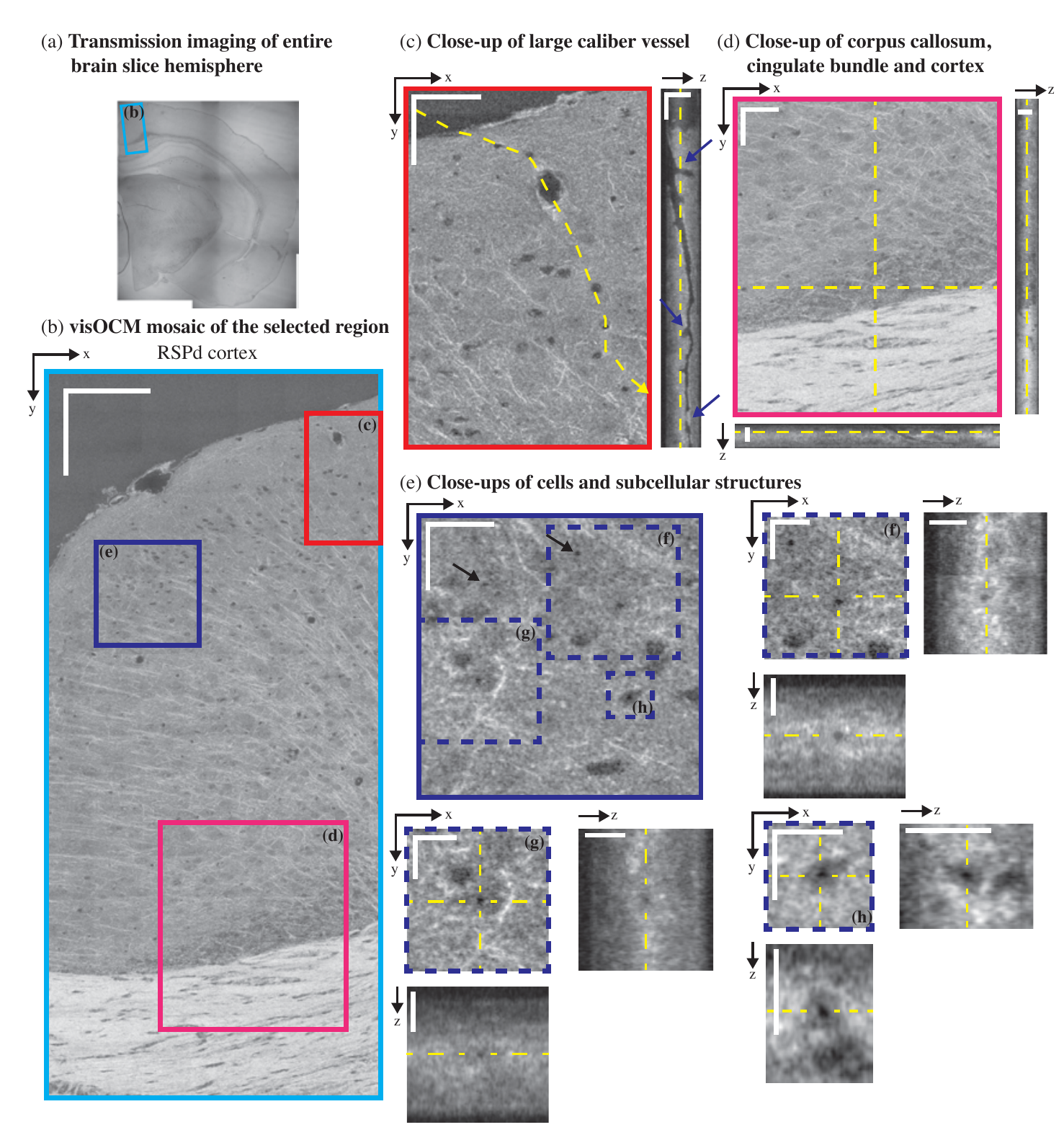}
\caption{\textit{ex-vivo} visOCM imaging of the RSPd cortex in a B6SJL/f1 mouse brain slice. (a) A transmission image of the entire half hemisphere was performed to locate part of the RSPd cortex (blue rectangle). A mosaic of the area of interest was then obtained with visOCM (b), where one can appreciate the presence of fibers, vessels and cells. A large penetrating vessel (c) can be observed through the difference in contrast between its hollow lumen and the back-scattering of the surrounding tissue. Examples of bifurcations and potential clogging of the vessel are pointed by arrowheads in the orthogonal view. Fibers appear as thin oriented bright structures and are present in the cortex and in the corpus callosum (d). Finally, sub-cellular features can also be observed as darker spots within the cell bodies, as shown in (e--h) and pointed by arrowheads in (e). Scalebars: 150 $\mu$m in (b), 50 $\mu$m in the en-face view of (c--d), 20 $\mu$m in the orthogonal views of (c--d) and in the en-face view of (e), 10 $\mu$m in the en-face and orthogonal views of (f--g).}
\label{Fig:RSPd}
\end{figure}

\subsection{Brain slice imaging}
The imaging performance of visOCM was first demonstrated by imaging cortical structures in fixed mice brain slices ($\sim$30 \textmu m thick) of both healthy and alzheimeric mice. 

The lateral and axial resolution and contrast offered by the visible spectrum allows resolving several different entities present in the brain such as myelinated fibers, vascular structures, cell bodies and amyloid plaques.

\subsubsection{Myelin fibers}

Similarly to previous studies performed with OCM at longer wavelengths \cite{Bolmont2012a,Srinivasan2012a,Leahy2013,Magnain2015}, myelin and neural fibers appear as bright linear structures through their increased back-scattering. In Figures \ref{Fig:PTLp}(a) and (c), fibers  emerge from the corpus callosum and the cingulum bundle and spread throughout the cortical column. The corpus callosum, shown in Figure \ref{Fig:RSPd}(d), contains a high density of fibers and can be distinguished as a bright region with orientated stripes. The cortex is characterized by a lower density of fibers with a higher variability in their orientation.

\subsubsection{Vascular structures}

Vascular compartments, from large calibres (penetrating arterioles and ascending venules) to the smallest capillaries can be discriminated from the background tissue as either hollow tubes (from the empty lumen) or as thin bright structures. Figure \ref{Fig:RSPd}(c) shows the en-face and orthogonal slice of a large penetrating vessel where one can distinguish the hollow lumen from the tissue and visualise bifurcations along the propagation of the vessel. Additionally, the edges of the vessel exhibit an increased signal, which could either be caused by a change in the scattering properties of the vessel's membrane or its surrounding tissue (for example vascular smooth muscle).
Smaller vessels, such as capillaries, can also be observed in the tomograms and appear either as dark or bright structures compared to the surrounding tissue. Figures \ref{Fig:PTLp}(e) and (f) show both dark and bright capillary structures and show that it is possible to trace their trajectory regardless of their contrast. Similarly to the large vessel in \ref{Fig:RSPd}(c), the dark contrast is indicative of a lack of scatterers within the lumen of the capillary. The bright contrast, on the other hand, could originate from scatterers filling the vessel's lumen (i.e. clogging during the perfusion procedure) or from the different scattering properties of the vessel's boundary.

\subsubsection{Cell bodies}

In addition to neuronal fibers and capillaries, visOCM imaging allows visualizing different cell body types through their different contrast with respect to the extracellular space. Figures \ref{Fig:PTLp}(b) and \ref{Fig:RSPd}(b) present mosaics covering parts of the posterior parietal association areas (PTLp1) and the retrosplineal area (RSPd) respectively, where one can observe two main cell body types with different contrasts and shapes. As highlighted in Figure \ref{Fig:PTLp}(d), some of the cells appear as dark spherical shapes, due to a decreased back-scattering. The second type of cells visible in Figure \ref{Fig:PTLp}(c) have similar contrast than the extracellular space and are larger. In addition to their different shapes and back-scattering properties, the two cell types also appear to be present in different regions of the cortical column: the darker and smaller cells are denser in the upper layers whereas the larger cells seem more prominent in the deeper layers, closest to the corpus callosum. Cells in the cornu ammonis area 1 (CA1), as shown in Figure \ref{Fig:5xFAD}(b), are also characterized by a darker contrast, similar to the cells in the upmost layers of PTLp1. Additionally, one can notice in the RSPd brain slice the presence of a small darker substructure within the body of certain cells. Figure \ref{Fig:RSPd}(e--h) displays a selection of these cells and their dark subcellular structure. Although a more complete study is necessary to elucidate the nature of this feature, our experience in live-cell imaging (results shown in Figure \ref{Fig:cells}) has shown that a similar contrast is present in what seems to be the nucleus. The orthogonal views and tile (g) of Figure \ref{Fig:PTLp} show that the signal acquired with visOCM extends throughout the depth of the tissue slice, although a loss in intensity and blurring can be observed in the deeper layers.

\subsubsection{Amyloid plaques}

\begin{figure}[htbp]
\centering\includegraphics[width=13cm]{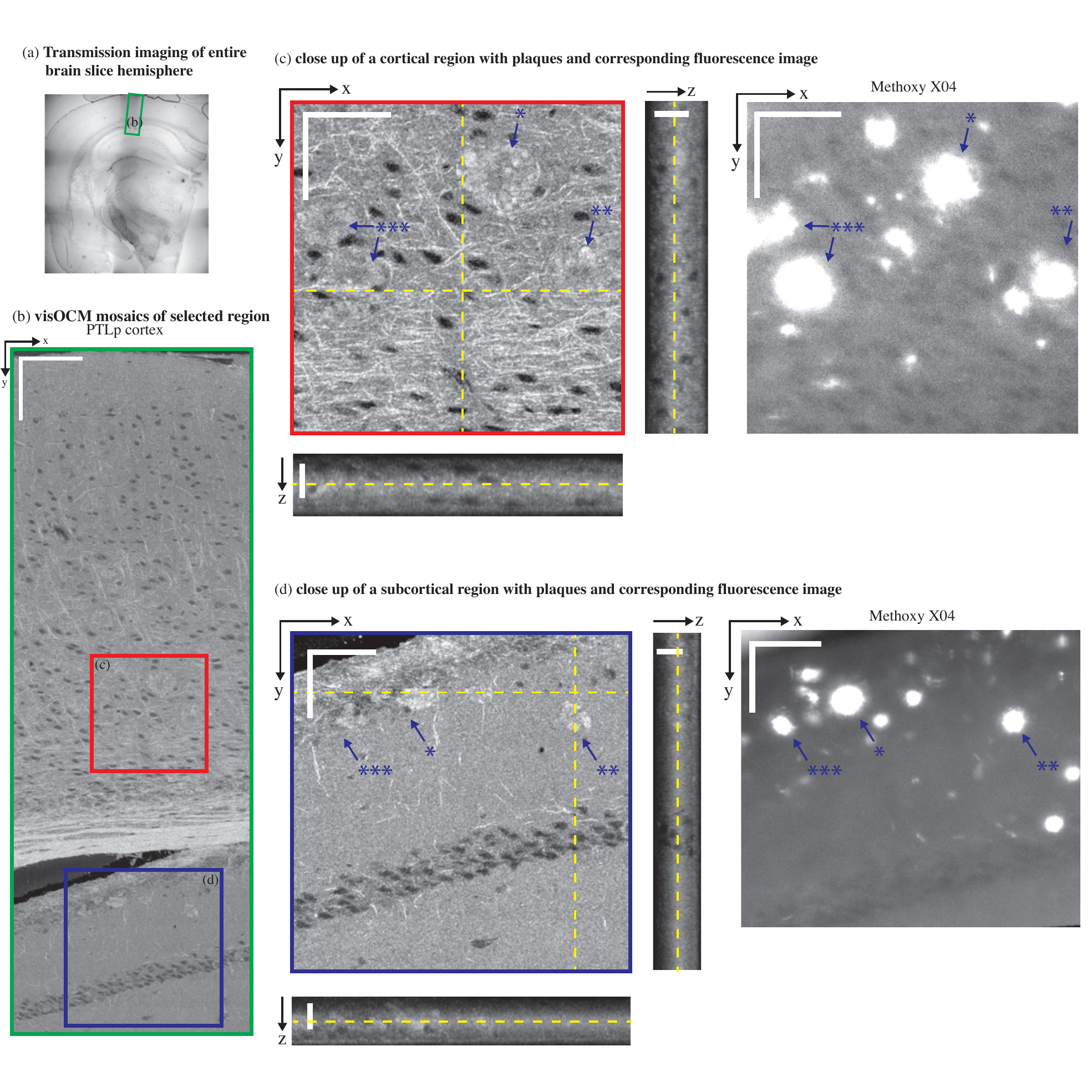}
\caption{\textit{ex-vivo} imaging of cortical and subcortical structures in a 5xFAD mouse brain slice. (a) A transmission image shows the location of the area where a visOCM image mosaic was obtained (b). The visOCM mosaic reveals fibers, cells and amyloid plaques in both cortical and subcortical structures. Close-ups of areas of interest containing amyloid plaques, in both cortical and subcortical regions, are displayed with their en-face views and corresponding fluorescence image (c--d). In the visOCM image, plaques can be seen as irregular high intensity regions.  Scalebars: 150 $\mu$m in (b), 50 $\mu$m in the en-face and 20 $\mu$m m in the orthogonal views of (c--d).}
\label{Fig:5xFAD}
\end{figure}

Previous work from our group has shown that amyloid plaques can be distinguished from cerebral tissue using xfOCM operating at 800 nm through their increased scattering \cite{Bolmont2012a}. In continuation of this work, we imaged brain slices of an alzheimeric mouse model with our novel visOCM system, with the expectation that the increased spatial resolution and different illumination wavelength would shed light on the details of these aggregates. We therefore imaged a part of the PTLp cortex and subcortical structures (CA1) of a 5xFAD mouse. As shown in Figure \ref{Fig:5xFAD}(b) and similarly to the results in Figures \ref{Fig:PTLp}(b) and \ref{Fig:RSPd}(b), the cortex is characterized by a high density of fibers and of cell bodies. The subcortical region, below the corpus callosum, has a slightly lower intensity compared to the cortex and presents a line of cells (CA1). Similarly to the results obtained at 800 nm \cite{Bolmont2012a}, the amyloid plaques manifest themselves as high intensity regions with a darker core. The increased spatial resolution of the system reveals with great detail the irregular shape of these aggregates, as shown in \ref{Fig:5xFAD}(c--d). The plaques are, as expected, present in cortical and also subcortical regions \cite{Jawhar2012}, where the slightly decreased intensity of the cerebral tissue provides a higher contrast between the plaques and the background. The location of the plaques was colocalized with fluorescence imaging of Methoxy-X04 using a commercial widefield microscope (Axiovert 200M, Zeiss), a 20x / 0.5 NA objective and the DAPI filter set (Excitation filter: 365 nm, dichroic mirror: 395 nm, Emission filter: 445/50 nm). As shown in tiles (c--d) of Figure \ref{Fig:5xFAD}, the locations of the aggregates in visOCM are in agreement with the location of the labelled structures in the fluorescence image.

\subsection{Cell imaging}

In addition to the imaging of tissue structures, we performed imaging of live macrophages in a cell culture. A mosaic of tomograms of these macrophages is shown in Figure \ref{Fig:cells}(a). The capabilities of the extended-focus can be appreciated in the orthogonal views of the tomogram, where the signal extends sufficiently in depth to reveal the three-dimensional organisation of the culture, with certain cells lying on top of other cells. The strongly scattering cytoplasm of the cell appears as a bright structure surrounding a darker subvolume, which most likely corresponds to the cell nuclei. The increased lateral resolution of the system allows resolving the filopodia on certain cells.

\subsection{Dynamic signal imaging}
\label{sec:DynSigIm}
\begin{figure}[htbp]
\centering\includegraphics[width=11cm]{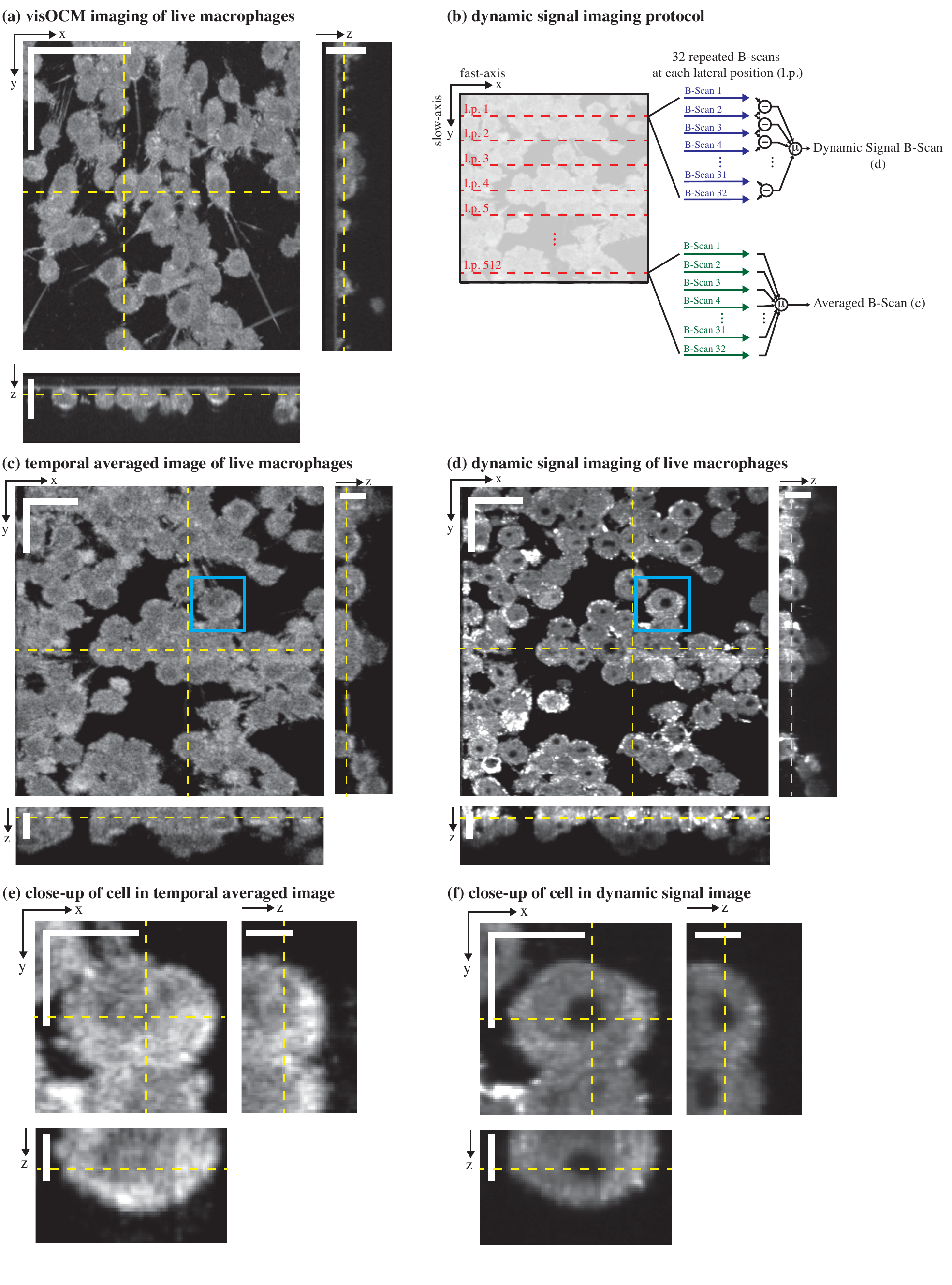}
\caption{visOCM imaging of murine macrophages. (a) A mosaic of macrophages obtained with visOCM with its orthogonal views reveals the three-dimensional organisation of cells in a culture. We further explored the capabilities of visOCM by applying a protocol similar to OCT angiography (b). The protocol entails imaging each lateral position along the slow axis 32 times (32 repeated B-scans per location). These 32 B-scans are either averaged or undergo a point-wise complex subtraction to obtain an averaged image (c) or a view of the dynamic components of the tomogram (d) respectively. The averaged image is identical in contrast to (a), whereas the dynamic signal image further reveals compartments within the cell, as either darker or brighter subregions (d) Close-ups of a selected cell in (c) and (d) are shown in (e) and (f) respectively, highlighting the differences in contrast between the averaged and dynamic images. Scalebars: 50 $\mu$m in the en-face view in (a), 20 $\mu$m in the orthogonal views in (a) and in the en-face views of (c--d), 10 $\mu$m in the orthogonal views in (c--d) and in the en-face views of (e--f), 5 $\mu$m in the orthogonal views in (e--f).}
\label{Fig:cells}
\end{figure}

In a second step, we analysed the dynamic properties of the scattering signal from living cells. In contrast to previous dynamic signal imaging techniques using the autocorrelation function or the standard deviation of the OCT signal \cite{Lee2013,Oldenburg2015,Apelian2016}, we extracted the dynamic component of the back-scattering through a point-wise subtraction of the complex OCM signal (Figure \ref{Fig:cells}(b)), as developed by Srinivasan et al. for OCT angiography \cite{Srinivasan2010a}. A time-series of scattering fluctuations per voxel was obtained by sampling each transverse position (B-scan) 32 times with a timestep of \textDelta t = 27 ms. Each timepoint was then temporally high-pass filtered (through a point-wise complex subtraction) and then averaged. A temporally averaged image was also obtained by averaging the repeated acquisitions.
The results of this operation are shown in Figure \ref{Fig:cells}(c) and Figure \ref{Fig:cells}(d) showing the averaged and dynamic signals respectively. Interestingly, in addition to the lower intensity of the cell nucleus already present in the static imaging, the dynamic imaging further reveals details within the cell body. As shown in Figure \ref{Fig:cells}(d), darker regions are present within the nuclei of the cells and brighter spots can be observed within the cytoplasm. Finally, the interface between the nuclei and the cytoplasm appears as a fine dark structure in certain cells. These differences are further revealed in Figure \ref{Fig:cells}(e) and (f) showing close-ups of a selected cell in the temporally averaged and dynamic signal image respectively.

\section{Conclusion}

In this work, we presented a novel OCM system, called visOCM, combining an extended-focus \cite{Leitgeb2006}, a dark-field detection \cite{Villiger2010}, a high-NA objective and an ultra-broad illumination spanning from the visible to the near-infrared wavelength range. As demonstrated here, the combination of these features provides an almost isotropic submicron resolution, maintained over > 40 \textmu m in depth. The capabilities of visOCM were demonstrated by imaging brain tissue slices of healthy and alzheimeric mice and macrophages cell cultures. The imaging of brain tissue with visOCM reveals several cortical structures as vessels, capillaries and cells. Interestingly, these different structures exhibited a wide range of contrasts, even within the same structure type. Capillaries could be observed as both dark or bright hollow structures. Although one cannot discard changes in tissue caused by the sample preparation (i.e. clotting and vessel collapsing due to the perfusion), the bright contrast could emanate from the presence of tissue bordering the vessels (such as smooth muscle cells or pericytes). Conversely, the dark contrast arises from a lack of scatterers from the hollow lumen of the vessel. Previous studies involving OCT and OCM imaging of brain tissue showed that certain cells could be identified through their different contrast within the cerebral tissue \cite{Lee2013,Srinivasan2012a,Magnain2015,Tamborski2016}. Srinivasan et al. and Tamborski et al. observed that certain neuron types could be discriminated from the tissue by their dark contrast using an OCM operating at 1300 nm and 800 nm respectively \cite{Srinivasan2012a,Tamborski2016}. By exploiting a higher lateral resolution and by shifting the illumination spectrum to the visible wavelength range, we show that this intrinsic contrast seems to vary between cell types and thus could be used in the future to identify different types of neurons and inter-neurons. Furthermore, the increased resolution of the visOCM system reveals subcellular features, potentially the cell nuclei. Future work will focus on elucidating the causes of these different contrasts and attempt to discriminate and potentially classify the cells according to their intrinsic scattering properties.

In a second step, we built upon a previous study performed at 800 nm by imaging brain slices of alzheimeric mice models with our novel system \cite{Bolmont2012a}. Similarly to our aforementioned results, the plaques appear as bright irregular structures in the tomograms obtained with visOCM. Although the exact nature of this particular contrast remains unknown, the presence of metals (such as Fe) in amyloid plaque cores, as described by Plascencia-Villa et al. \cite{Plascencia-Villa2016}, might provide a first hint into the cause of this phenomenon. In fact, a previous study from our group showed that the contrast of the Langerhans islets in OCM images originated from the presence of Zn crystals \cite{Berclaz2016}. Alternatively, this contrast could also arise from polarization effects as the plaques have been shown to have birefringent properties \cite{Baumann2017}. 

Finally, we demonstrated the performance of visOCM by imaging live-cells in cultures. The extended-focus and high acquisition rates available with the platform allows fast imaging of the three-dimensional structure of cell cultures. The shift in illumination wavelength and increase in NA provides the resolution necessary to identify thin structures such as filopodias and sufficient contrast to visualize seemingly sub-cellular structures.
Recent studies have highlighted the importance of studying the dynamic properties of cellular back-scattering to understand cellular function \cite{Leroux2016}. In this context, we explored the possibility to perform dynamic imaging, as already performed in full field OCM \cite{Apelian2016}, OCM \cite{Sison2017} and phase imaging \cite{Ma2016}, with our novel imaging platform to further discriminate between subcellular compartments. We applied a protocol devised for OCT angiography \cite{Srinivasan2010a} and show that a \textit{dynamic} contrast can be obtained in a culture of macrophages. Although more work is needed to identify the different regions and the nature of the signals, one can appreciate the increased contrast provided by the protocol. Overall, the combination of the extended-focus, the high isotropic resolution and high acquisition rates make visOCM an ideal platform to monitor fast processes occurring at the subcellular level in cell cultures.

In addition to our demonstration of the capabilities of our novel visOCM system, we have introduced a strategy for physical dispersion compensation. We used a multivariate linear regression to model the OPD of the system prior to alignment and present a processing pipeline and a hardware unit to finely minimize the dispersion during the construction of the microscope. In this work, we opted for a hardware dispersion compensation strategy, however numerical dispersion compensation techniques could also have been used and will be explored in future work.

Ultimately, visOCM offers label-free 3D imaging of tissue and cell structures but remains limited through its lack in specificity. Our results show that it is possible to discriminate between structures in tissue and cells through their individual back-scattering properties and potentially through their dynamic signatures (as revealed by our dynamic imaging). Nevertheless, these contrast mechanisms are limited and do not always provide the desired molecular specificity. In such regards, the visOCM platform could be augmented with a collinear fluorescence channel. This configuration would allow molecular-specific fluorescent imaging to be complemented with visOCM imaging, providing information on the overall structural context of the organism under investigation. 
In addition to fluorescence, the visible spectrum could be used to explore the spectroscopic signatures of specific molecules. Spectroscopic OCT has shown great promise in its ability to provide additional contrast mechanisms \cite{Morgner2000,Robles2011}. Having an illumination in the visible wavelength range could be used to visualize and discriminate between different endogeneous and/or exogenous contrast agents within the imaged tissue or cells.

\section*{Funding}
This study was partially supported by the Swiss National Science
Foundation (205321L\_135353 and 205320L\_150191), the Commission for Technology and Innovation (13964.1 PFLS-LS and 17537.2 PFLS-LS) and by the EU Framework Programme for Research and Innovation (686271).

\section*{Acknowledgments}
We thank Kristin Grussmayer for preparing the cells and B. Deplancke (LSBG, EPFL) for kindly giving us the murine macrophages cell line RAW 264.7.

\end{document}